\documentclass[journal]{IEEEtran}
\ifCLASSINFOpdf
\else
\fi
\usepackage{url}

\usepackage[numbers,sort&compress]{natbib}

\usepackage{stfloats}
\usepackage{graphicx}
\usepackage{amsmath}
\usepackage{amsthm}
\usepackage{multirow}
\usepackage{amssymb}
\usepackage{subfigure}
\newcommand*{\QEDA}{\hfill\ensuremath{\blacksquare}}
\usepackage{booktabs}

\usepackage{fancyhdr}
\usepackage{color}

\newtheorem{Remark}{Remark}
\newtheorem{Theorem}{Theorem}
\begin{document}
%
\title{Uplink Achievable Rate for Massive MIMO Systems with Low-Resolution ADC}
%
%
%
\author{Li Fan, Shi Jin,~\IEEEmembership{Member,~IEEE}, Chao-Kai Wen,~\IEEEmembership{Member,~IEEE} and Haixia Zhang,~\IEEEmembership{Senior Member,~IEEE}
\thanks{Manuscript received July 11, 2015; accepted October 12, 2015. The work of L. Fan and S. Jin was supported by the National Natural Science Foundation of China under  Grants 61531011, 61450110445, and 61222102, the  Natural Science Foundation of Jiangsu Province under Grant BK2012021, the  International Science and Technology Cooperation Program of China under  Grant 2014DFT10300. The work of C.-K. Wen was supported by the ITRI in Hsinchu, Taiwan, under Grant E352J33120. The work of H. Zhang was supported by the National Natural Science Foundation of China under  Grant 61371109.}
\thanks{L. Fan and S. Jin are with the National Mobile Communications Research
Laboratory, Southeast University, Nanjing 210096, China (e-mail: ff502120@163.com; jinshi@seu.edu.cn).}
\thanks{C.-K. Wen is with the Institute of Communications Engineering, National
Sun Yat-sen University, Kaohsiung 80424, Taiwan (e-mail: ckwen@ieee.org).}
\thanks{Haixia Zhang is with Wireless Mobile Communication and Transmission Laboratory, Shandong University, China (haixia.zhang@sdu.edu.cn).}}
\maketitle
\thispagestyle{fancy}
\begin{abstract}
In this letter, we derive an approximate analytical expression for the uplink achievable rate of a massive multi-input multi-output (MIMO) antenna system when
finite precision analog-digital converters (ADCs) and the common maximal-ratio combining technique are used at the receivers. To obtain this expression, we treat
quantization noise as an additive quantization noise model. Considering the obtained expression, we show that low-resolution ADCs lead to a decrease in the
achievable rate but the performance loss can be compensated by increasing the number of receiving antennas. In addition, we investigate the relation between the
number of antennas and the ADC resolution, as well as the power-scaling law. These discussions support the feasibility of equipping highly economical ADCs with low
resolution in practical massive MIMO systems.
\end{abstract}


%
\IEEEpeerreviewmaketitle

\section{Introduction}
%
%
%
%
\IEEEPARstart{M}{assive} multi-input multi-output (MIMO) antenna technology has emerged as an attractive candidate technique for 5G mobile networks because of its
potential to significantly improve the capacity of wireless communication systems \cite{Ref 1}. Such a technique mainly uses hundreds of antennas at the base
station (BS) to serve a relatively small number of user terminals on the same time-frequency channel \cite{Ref 1,Ref 2,Ref 3}. Research \cite{Ref 2} showed that
simple linear processing such as maximal-ratio combining (MRC) in massive MIMO systems can achieve very high spectral efficiency. In addition, massive MIMO has
been proven able to achieve a significant gain in energy efficiency \cite{Ref 3}.

The large number of antennas required in massive MIMO causes a substantial increase in hardware cost and power consumption. This fact motives studies on quantized
MIMO systems \cite{Ref 4,Ref 5,Ref 6}, where each receiver antenna uses a very low resolution (e.g., $1-3$ bits) analog-to-digital converter (ADC). The capacity of
$1$-bit quantized MIMO was studied in \cite{Ref 4}, where the exact nonlinearity property of a quantizer is considered. Rather than using the nonlinear quantizer
model, studies \cite{Ref 5,Ref 6} treat the quantization noise as additive and independent noise, as encapsulated by the additive quantization noise model (AQNM).
Although the model is an approximation, AQNM is widely used because it facilitates analysis and provides insights into quantized systems. Under AQNM, the effect of
ADC resolution and bandwidth on the achievable rate were investigated in \cite{Ref 6}.

In this letter, AQNM is used in investigating the uplink achievable rate affected by quantizated massive MIMO systems. In particular, we consider an MRC receiver
that can be performed in a non-centralized manner to avoid the unscalable data transport overhead required by other existing receivers (e.g., zero-forcing
receiver). Under the assumption of perfect channel state information (CSI) at BS, a tight approximate expression for an achievable uplink rate that holds for
arbitrary number of antennas is provided. The approximate expression agrees with previous works in \cite{Ref 3} and \cite{Ref 7} as the quantizers have infinite
precision. On the basis of the derived expression, the relationship between the number of antennas and the ADC resolution, as well as the power-scaling law, is
investigated.





\section{System Model}
Consider the uplink of a multi-user MIMO (MU-MIMO) system formed by a BS equipped with an array of $M$ antennas and serving $N$ single antenna user terminals in
the same time-frequency resource. The received $M$ dimensional vector ${\bf{y}}$ at BS can be expressed as \cite{Ref 2}
\begin{equation}
\label{eq 1}
{\bf{y}} = \sqrt {{p_u}} {\bf{Gx}} + {\bf{n}},
\end{equation}
where $\bf{G}$ represents the $M \times N$ channel matrix between BS and users, $ {\bf{x}}$ denotes the $N \times 1$ vector of symbols transmitted by all users,
${p_u}$ is the average transmitted power of each user, and ${ {\bf{n}} \sim {\cal C}{\cal N}\left( {0,{\bf{I}}} \right) }$ is the additive white Gaussian noise
vector.

We denote the independent channel coefficient between the $n$th user and the $m$th antenna at the BS as ${g_{mn}} = {\left[ {\bf{G}} \right]_{mn}}$, which models
independent fast fading, geometric attenuation, and log-normal shadow fading \cite{Ref 2}. The coefficient ${g_{mn}}$ can be written as
\begin{equation}
\label{eq 2}
{g_{mn}} = {h_{mn}}\sqrt {{\beta _n}},
\end{equation}
where $h_{mn}$ is the fast-fading coefficient from the $n$th user to the $m$th antenna of BS, and ${\beta _n}$ models both geometric attenuation and shadow fading,
which is assumed to be constant across the antenna array. In matrix form, we write
\begin{equation}
\label{eq 3}
{\bf{G}} = {\bf{H}}{{\bf{D}}^{{1 \mathord{\left/
 {\vphantom {1 2}} \right.
 \kern-\nulldelimiterspace} 2}}},
\end{equation}
where ${{\bf H} = [h_{mn}] }$ is the ${M \times N}$ fast fading matrix between the users and BS, and $\bf{D}$ is the ${N \times N}$ diagonal matrix with diagonal entries $\{ \beta _n \}$ .


Assuming the gain of the automatic gain control is set appropriately, we use AQNM and formulate the quantizer outputs as \cite{Ref 6}
\begin{equation}
\label{eq 5}
{{\bf{y}}_{\sf q}} = \alpha {\bf{y}} + {{\bf{n}}_{\sf q}} = \alpha \sqrt {{p_u}} {\bf{Gx}} + \alpha {\bf{n}} + {{\bf{n}}_{\sf q}}
\end{equation}
with ${\alpha =1 - \rho} $, where $\rho$ is the inverse of the signal-to-quantization-noise ratio, and ${{\bf{n}}_{\sf q}}$ is the additive Gaussian quantization noise vector that is uncorrelated with $\bf{y}$. Let $b$ be the number of quantization bins. We assume that the input to the quantizer is Gaussian.
Accordingly, for the non-uniform scalar minimum mean-square-error quantizer of a Gaussian random variable, the values of $\rho$ are listed in Table \ref{Tab 1} for
$b \le 5$ and can be approximated by $\rho  = {\textstyle{{\pi \sqrt 3 } \over 2}} \cdot {2^{{\rm{ - }}2b}}$ for ${b > 5} $.
\begin{table}
\centering
    \caption{\label{Tab 1}$\rho$ for different ADC quantization bits $b$}
    \begin{tabular}{lccccc}
        \toprule
        $b$  & 1 &2 & 3 & 4 & 5 \\
        \midrule
        $\rho$ &0.3634&0.1175&0.03454&0.009497&0.002499\\
        \bottomrule
    \end{tabular}
\end{table}

For a fixed channel realization $\bf{G}$, the covariance of ${{\bf{n}}_{\sf q}}$ is given by
 \begin{align}
 \label{eq 6}
{{\bf{R}}_{{{\bf{n}}_{\sf q}}{{\bf{n}}_{\sf q}}}}
 &= {\mathbb E}\!\left\{ {{{\bf{n}}_{\sf q}}{\bf{n}}_{\sf q}^H \big| {\bf{G}}} \right\} \notag \\
 &= \alpha \left( {1 - \alpha } \right){\sf diag}{\left( {p_u{\bf{G}}{{\bf{R}}_{x}}{{\bf{G}}^H} + {\bf{I}}} \right)},
\end{align}
where ${{\bf{R}}_{x}} $ is the input covariance matrix . Since ${{\bf{R}}_{x}} = {\bf{I}}$, (\ref{eq 6}) can be expressed as
  \begin{equation}
 \label{eq 7}
{{\bf{R}}_{{{\bf{n}}_{\sf q}}{{\bf{n}}_{\sf q}}}} = \alpha \left( {1 - \alpha } \right){\sf diag}{\left( {p_u{\bf{G}}{{\bf{G}}^H} + {\bf{I}}} \right)}.
\end{equation}

\section{Analysis of Achievable Uplink Rate}
In this section, we derive a tractable approximate expression of the achievable uplink rate for the MRC receiver. Using the expression, we analyze the effect of
ADC resolution on the achievable uplink rate. For the MRC receiver, the quantized signal vector is processed as
\begin{equation}
\label{eq 8}
{\bf{r}} = {{\bf{G}}^H}{{\bf{y}}_{\sf q}}.
\end{equation}
Substituting (\ref{eq 5}) into (\ref{eq 8}) results in
\begin{equation}
\label{eq 9}
{\bf{r}} = \sqrt {{p_u}} \alpha {{\bf{G}}^H}{\bf{Gx}} + \alpha {{\bf{G}}^H}{\bf{n}} +
{{\bf{G}}^H}{{\bf{n}}_{\sf q}}.
\end{equation}
From (\ref{eq 9}), the $n$th element of $\bf{r}$ can be expressed as
\begin{equation}
\label{eq 10}
{r_n} = \sqrt {{p_u}} \alpha {\bf{g}}_n^H{{\bf{g}}_n}{x_n} + \sqrt {{p_u}} \alpha \sum\limits_{i = 1,i \ne n}^N {{\bf{g}}_n^H{{\bf{g}}_i}{x_i}}  + \alpha {\bf{g}}_n^H{\bf{n}}\; + {\bf{g}}_n^H{{\bf{n}}_{\sf q}},
\end{equation}
where ${{\bf{g}}_n}$ is the $n$th column of $\bf{G}$. Given a channel realization ${\bf{G}}$, the noise-plus-interference term is a random variable with zero mean
and variance
\begin{multline} \label{eq:IfVar}
{\cal I}_{{\bf G}} = {p_u}{\alpha}^2 \sum\limits_{i = 1,i \ne n}^N {{{\left| {{\bf{g}}_n^H{{\bf{g}}_i}} \right|}^2}}  + {\alpha}^2 {{\left\| {{{\bf{g}}_n}} \right\|}^2} \\
 + \alpha\left( {1 - \alpha } \right){\bf{g}}_n^H{\sf diag}\left( {{p_u}{\bf{G}}{{\bf{G}}^H} + {\bf{I}}} \right){{\bf{g}}_n}.
\end{multline}
Following the assumption in \cite{Ref 9}, we model this term as additive Gaussian noise independent of ${x_n}$ and derive the ergodic achievable uplink rate of the $n$th user as
\begin{equation}
\label{eq 11}
{R_n} = {\mathbb E}\!\left\{ {{{\log }_2}\left( {1 + \frac{{{p_u}{\alpha ^2}{{\left\| {{{\bf{g}}_n}} \right\|}^4}}}{ {\cal I}_{{\bf G}} }} \right)} \right\},
\end{equation}
where ${\cal I}_{{\bf G}}$ is given in (\ref{eq:IfVar}), and  the expectation is taken over the fast-fading coefficients $\{ h_{mn} \}$. No efficient way is able to directly calculate the achievable rate in (\ref{eq 11}). Therefore, we derive
an approximation that is presented as follows:

\begin{Theorem}
Using MRC receivers with perfect CSI in the quantized MIMO systems, we can approximate the achievable uplink rate of the $n$th user by
\begin{equation}
\label{eq 13}
{{\widetilde R}_n} = {\log _2}{\left( {1 + \frac{{{p_u}{\alpha}{\beta _n}\left( {M + 1} \right)}}{{\cal I}}} \right)}
\end{equation}
where
\begin{equation}
 {\cal I} = {{p_u}\alpha \sum\limits_{i = 1,i \ne n}^N {{\beta _i}}  + {p_u}\left( {1 - \alpha } \right){\left( {\sum\limits_{i = 1}^N {{\beta _i}}  + {\beta _n}} \right)} + 1}.
\end{equation}
\end{Theorem}
\emph{Proof:} Applying \cite[Lemma~1]{Ref 7}\footnote{This lemma has been proved as a tight approximation in \cite{Ref 7} and has also been used by \cite{Ref
10}.}, we begin by approximating $R_n$ with
\begin{equation}
\label{eq 14}
{{\widetilde R}_n} = {\log _2}{\left( {1 + \frac{{{p_u}{\alpha ^2}{\mathbb E}{\left\{ {{{\| {{{\bf{g}}_n}} \|}^4}} \right\}}}}
 { {\mathbb E}\{ {\cal I}_{{\bf G}} \} }} \right)}
\end{equation}
where
\begin{multline} \label{eq:14-1}
 {\mathbb E}\{ {\cal I}_{{\bf G}} \} = {p_u}\alpha^2 \sum\limits_{i = 1,i \ne n}^N {{\mathbb E}\Big\{ {{{\left| {{\bf{g}}_n^H{{\bf{g}}_i}} \right|}^2}} \Big\}}
  + \alpha^2 {\mathbb E}\Big\{ {{{\left\| {{{\bf{g}}_n}} \right\|}^2}} \Big\} \\ + \alpha\left( {1 - \alpha } \right) {\mathbb E}\Big\{ {{\bf{g}}_n^H{\sf diag}\left( {{p_u}{\bf{G}}{{\bf{G}}^H} + {\bf{I}}} \right){{\bf{g}}_n}} \Big\} .
\end{multline}
To obtain a tractable expression from (\ref{eq 14}), we have to calculate the expectation of several terms. {{Recall ${{\bf{g}}_n} = \sqrt {{\beta _n}} {{\bf{h}}_n}$
and ${{h_{mn}} \sim {\cal C}{\cal N}\left( {0,1} \right)}$. Then, the computation of the expectation of squared norm terms ${\left\| {{{\bf{g}}_n}} \right\|^2}$
yields
\begin{equation}
\label{eq 15}
{\mathbb E}{\left\{ {{{\left\| {{{\bf{g}}_n}} \right\|}^2}} \right\}} = \beta_nM.
\end{equation}
This equation holds for the condition where the random variable ${{{\left\| {{{\bf{g}}_n}} \right\|}^2}} $ is gamma-distributed with shape $M$ and scale $\beta_n$ and can be denoted by ${{{\left\| {{{\bf{g}}_n}} \right\|}^2}}  \sim \Gamma \left( {M,\beta_n} \right)$. Then, we can obtain the variance of ${{{\left\| {{{\bf{g}}_n}} \right\|}^2}} $:
\begin{equation}
\label{eq 16}
{{\bf{V}}\!{\rm{ar}}}{\left\{ {{{\left\| {\bf{g}}_n \right\|}^2}} \right\}} = \beta _n^2M.
\end{equation}

Subsequently, we compute the expectation of squared norm terms ${\left| {{\bf{g}}_n^H{{\bf{g}}_i}} \right|^2}$. First, in the case of $i = n$, we have
\begin{equation}
\label{eq 17}
{\mathbb E}\!\left\{ {{{\left\| {{{\bf{g}}_n}} \right\|}^4}} \right\} ={{\bf{V}}\!{\rm{ar}}}{\left\{ {{{\left\| {\bf{g}}_n \right\|}^2}} \right\}}+{\left({\mathbb E}\!\left\{ {{{\left\| {{{\bf{g}}_n}} \right\|}^2}} \right\}\right)}^2.
\end{equation}
Applying (\ref{eq 15}) and (\ref{eq 16}) to (\ref{eq 17}), we obtain
 \begin{equation}
\label{eq 19}
{\mathbb E}\!\left\{ {{{\left\| {{{\bf{g}}_n}} \right\|}^4}} \right\} = \beta^2_n({M^2} + M).
\end{equation}
Second, we consider the case of $i \ne n$, where many uncorrelated items exist. In this case, we have
\begin{align}
\label{eq 20}
{\mathbb E}\!\left\{ {{{\left| {{\bf{g}}_n^H{{\bf{g}}_i}} \right|}^2}} \right\}= \beta_n\beta_iM.
\end{align}

Then, we compute the last term in (\ref{eq:14-1}). The $m$th diagonal element of ${\sf diag}\left( {{p_u}{\bf{G}}{{\bf{G}}^H} + {\bf{I}}} \right)$ can be expressed as
\begin{equation}
\label{new eq 22}
{\left[ {{\sf diag}\left( {{p_u}{\bf{G}}{{\bf{G}}^H} + {\bf{I}}} \right)} \right]_{mm}} = 1 + {p_u}\sum\limits_{i = 1}^N {{{\left| {{g_{mi}}} \right|}^2}}.
\end{equation}
Employing (\ref{new eq 22}) to expand the last term in (\ref{eq:14-1}), we obtain
\begin{align}
   & {\mathbb E}{\left\{ {{\bf{g}}_n^H{\sf diag}\left( {{p_u}{\bf{G}}{{\bf{G}}^H} + {\bf{I}}} \right){{\bf{g}}_n}} \right\}}
  \notag \\
 = &\, {\mathbb E}{\left\{ {\sum_{m = 1}^M { {{{\left| {{g_{mn}}} \right|}^2}\Bigg( {1 + {p_u} \sum_{i = 1,i\not= n}^N {{{\left| {{g_{mi}}} \right|}^2}} +{p_u}\left|g_{mn}\right|^{2}} \Bigg)} } } \right\}}
   \notag \\
 = &\,  \sum_{m = 1}^M{\mathbb E}{\left\{\left|{g_{mn}}\right|^2\right\}} +{p_u}\sum_{m = 1}^M {\sum_{i = 1,i\not=n}^N {{\mathbb E}{\left\{ {{{\left| {{g_{mn}}} \right|}^2}} \right\}}{\mathbb E}\!\left\{ {{{\left| {{g_{mi}}} \right|}^2}}\right\}}}
  \notag \\
  &\, +{p_u}\sum_{m = 1}^M {\mathbb E}{\left\{ {{{\left| {{g_{mn}}} \right|}^4}} \right\}}. \label{eq 25}
\end{align}
An application of ${\mathbb E} \{ |g_{mn}|^2 \} = {\beta _n}$ and ${\mathbb E} \{ |g_{mn}|^4 \} = 2\beta _n^2$ to (\ref{eq 25}) yields
\begin{equation}
\label{eq 26}
{\mathbb E}{\left\{ {{\bf{g}}_n^H{\sf diag}\left( {{p_u}{\bf{G}}{{\bf{G}}^H}+{\bf{I}}} \right){{\bf{g}}_n}} \right\}} = M{\left( {{\beta _n}{\rm{ + }}{p_u}{\beta _n}\sum\limits_{i = 1}^N {{\beta _i}}  + {p_u}\beta _n^2} \right)}.
\end{equation}
Substituting (\ref{eq 15}), (\ref{eq 19}), (\ref{eq 20}), and (\ref{eq 26}) into (\ref{eq 14}) and simplifying the equation, we finally obtain the desired result in (\ref{eq
13}). \QEDA

Theorem 1 reveals the effect of the number of antennas $M$, number of quantization bits $b$, and transmit power $p_{u}$ on the rate performance. {{In contrast to \cite{Ref 3} and \cite{Ref 7}, Theorem 1 furthur involves the effect of ADC and embraces \cite{Ref 3} and \cite{Ref 7} as special cases.}} To ensure full and profound understanding of Theorem 1, we provide the following asymptotic results.
\par

\begin{Remark}
With fixed ${p_u}$ and $M$, when $b \to \infty $, (\ref{eq 13}) reduces to
\begin{equation}
\label{eq 27}
{{\widetilde R}_n} \to {\log _2}{\left( {1 + \frac{{{p_u}{\beta _n}\left( {M + 1} \right)}}{{{p_u}\sum_{i = 1,i \ne n}^N {{\beta _i}}  + 1}}} \right)}
\end{equation}
which agrees with a result derived previously in \cite{Ref 7} for the infinite precision case. This conclusion is reasonable because $b \to \infty $ means that the
quantization error brought by ADC can be ignored. Comparing (\ref{eq 27}) and (\ref{eq 13}), we notice that the
quantization effect $\alpha$ influences both the numerator and the denominator of (\ref{eq 13}). Thus, the result
of (\ref{eq 13}) cannot be obtained by fully modeling the quantization noise as an increased noise, where the quantization affects only the denominator.
\end{Remark}

\begin{Remark}
{
With fixed $b$ and $M$, when ${{p_u} \to \infty }$, (\ref{eq 13}) converges to
\begin{equation}
\label{eq 28}
{{\widetilde R}_n} \to {\log _2}{\left( {1 + \frac{{{\alpha}{\beta _n}\left( {M + 1} \right)}}{{\sum_{i = 1,i \ne n}^N {{\beta _i}} }+2(1-\alpha){\beta_n}}} \right)}.
\end{equation}
(\ref{eq 28}) shows that when the transmit power ${p_u}$ tends to infinity, the approximate rate approaches a constant that is dependent of the quantization bits. This observation indicates that the uplink rate performance degradation caused by the ADCs cannot be compensated by increasing the transmit power. Furthermore, (\ref{eq 28}) reveals that the uplink rate performance cannot be improved infinitely merely by increasing the transmit power of each user.  The reason is that both the desired signal power and the interference power caused by other users increase along with ${p_u}$.
}
\end{Remark}

\begin{Remark} Assume that the transmit power of each user is scaled with $M$ according to ${p_u} = {{{E_u}} \mathord{\left/{\vphantom {{{E_u}} M}}
\right.\kern-\nulldelimiterspace} M}$, where ${E_u}$ is fixed. Then (\ref{eq 13}) tends to
\begin{equation}
\label{eq 29}
{{\widetilde R}_n} \to {\log _2}\left( {1 + \alpha {\beta _n}{E_u}} \right),\mbox{ as } M \to \infty.
\end{equation}
 Note that $\alpha \geq 0$ increases with $b$ and
is upper bounded by 1, which indicates that the rate performance can be improved by increasing the number of quantization bits. Considering the case with infinite resolution, (\ref{eq 29}) becomes ${{\widetilde R}_n} \to {\log _2}\left( {1 + {\beta _n}{E_u}} \right)$, which aligns with the conclusion in \cite{Ref 3}. Comparing with the achievable rate in this case, (\ref{eq 29}) implies that even when low-resolution ADCs are used at the receivers, if the number of antenna $M$ can grow without limit, we can scale down the transmit power proportionally to ${1/M}$ to maintain the rate of a single-input single-output system with the transmit power ${\alpha E_u}$.
\end{Remark}

\section{Numerical Results}

In our simulation, we consider a hexagonal cell with radius of 1000 meters. The users are distributed randomly and uniformly over the cell, with the exclusion of a central disk radius ${r_h} = 100$ meters. The large-scale fading is modeled as ${\beta _n} = {{{z_n}} \mathord{\left/ {\vphantom {{{z_n}} {{{\left( {{{{r_n}} \mathord{\left/ {\vphantom {{{r_n}} {{r_h}}}} \right. \kern-\nulldelimiterspace} {{r_h}}}} \right)}^v}}}} \right. \kern-\nulldelimiterspace} {{{\left( {{{{r_n}} \mathord{\left/ {\vphantom {{{r_n}} {{r_h}}}} \right. \kern-\nulldelimiterspace} {{r_h}}}} \right)}^v}}}$, where $z_n$ is a log-normal variable with standard deviation  ${\sigma _{\rm shadow}}$, $r_{n}$ is the distance between the $n$th user and BS, and $v$ is the path loss exponent. Note that $\beta_n$ is fixed once the $n$th user is dropped in the cell, which agrees with the setting of the analytical derivation in Section III that $\beta_n$ is fixed, and the expectation is taken over the fast-fading coefficients. In all examples, we assume that ${\sigma _{\rm shadow}} = 8\;\mbox{dB}$, $v = 3.8$, and the small-scale fading follows Rayleigh distribution. We define the uplink sum rate of the entire system as $R = \sum\nolimits_{n = 1}^N {{R_n}} $.

We first conduct an experiment to validate the {accuracy~of} our proposed rate approximation in Theorem 1. In {Fig.~\ref{fig1}}, the simulated uplink achievable rate in (\ref{eq 11}) is compared with its corresponding analytical approximation in (\ref{eq 13}). In this example, $N = 10$ users have transmit power ${p_u} = 10\;\mbox{dB}$. Results are presented for three different quantizers with $1$, $2$, and $\infty$ bits, respectively. In all cases, a precise agreement between the simulation results and our analytical results can be found.

Then, we investigate the power-scaling law in (\ref{eq 29}). In this case, we choose ${{E_u} = 20\;\mbox{dB}}$, and $b$ has three different values 1, 2, and $\infty $. With ${p_u} = {{{E_u}} \mathord{\left/ {\vphantom {{{E_u}} M}} \right. \kern-\nulldelimiterspace} M}$, when $M$ increases, the results in Fig. \ref{fig2} show that the analytic approximations are consistent with exact values. As the figure shows, these curves eventually saturate with an increased $M$ which means that a balance exits between the increase and decrease of the sum rate caused by the increased $M$ and scaled-down ${p_u}$. As expected from the analysis of (\ref{eq 29}), the sum rates increase with $b$. However, the gaps between these curves narrow down with the increase of $b$. This finding implies that we do not need to equip BS with very high-resolution quantizers because the performance, which can be improved by increasing the number of quantization bits, is extremely limited. All of these observations agree with (\ref{eq 29}).

\begin{figure}
\includegraphics[width=8.7cm,height=6.6cm]{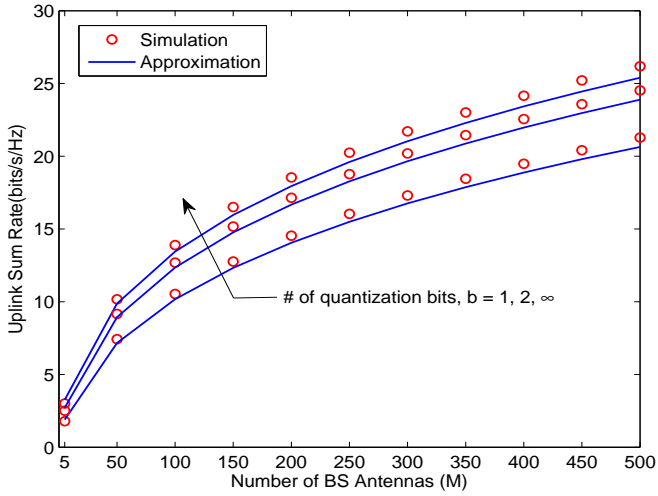}
\caption{\small{Uplink sum rate per cell versus the number of BS antennas, where $N = 10$ users have transmit power ${p_u} = 10$ dB.}}
\label{fig1}
\end{figure}

\begin{figure}
\includegraphics[width=8.7cm, height=6.6cm]{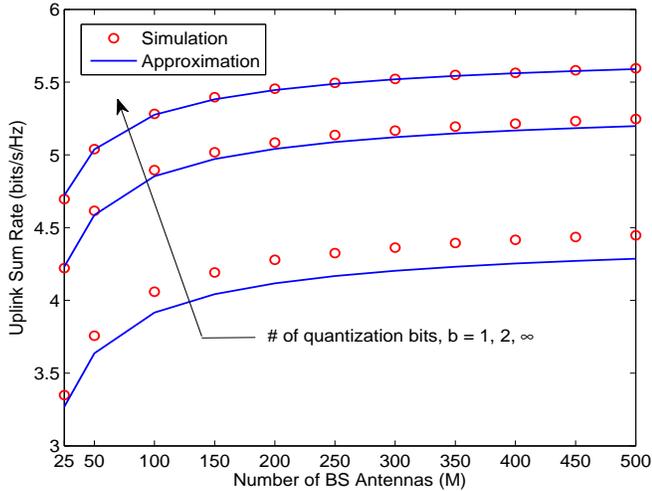}
\caption{\small{Uplink sum rate per cell versus the number of BS antennas, where $N = 10$ users have scaled-down transmit power ${p_u} = \frac{E_u}{M} $ dB. }}
\label{fig2}
\end{figure}
Finally, we illustrate energy efficiency, another important performance metric, which is defined as \cite{Ref BaiQ} $\eta  \triangleq BR/P$ with $P = {c_0}M{2^b} + {c_1}$, where ${c_0 = 10^{-4} }$ Watt and ${ {c_1} = 0.02 }$ Watt, $B$ is the bandwidth set to 1 MHz, and $R$ is the sum rate. Given the tightness between the simulated values and approximations, we adopt the approximation in (\ref{eq 13}) for analysis. In Fig. \ref{fig3}, we consider the same setting as in Fig. \ref{fig1}, but the number of BS antennas is fixed at $M = 100$. We observe that the sum rate converges to a fixed value obtained by (\ref{eq 27}). Moreover, the growth rate slows down with the increase of $b$ so that this increase also leads to significant degeneration of energy efficiency. Therefore, adopting low-resolution ADCs (e.g., $1-2$ bits) is a promising solution to ensure energy efficiency in massive MIMO systems.
\begin{figure}
\includegraphics[width=8.7cm, height=6.6cm]{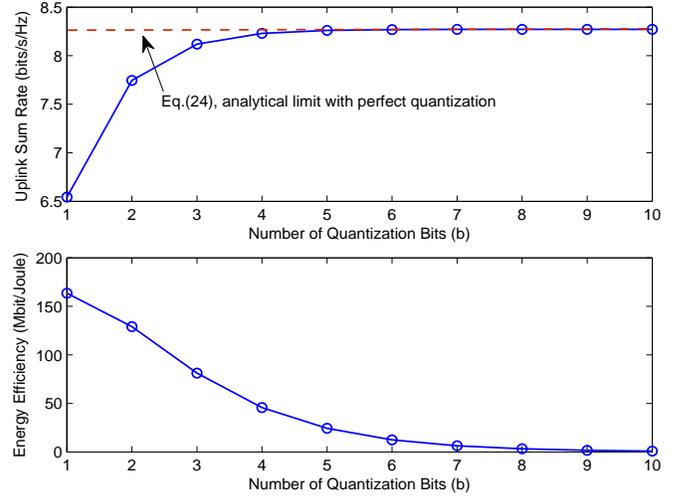}
\caption{\small{Uplink sum rate per cell and energy efficiency versus the number of quantization bits, where $N = 10$ users have transmit power ${p_u} = 10$ dB.}}
\label{fig3}
\end{figure}

\section{Conclusions}
We have derived a tight approximate expression for an achievable uplink rate in massive MIMO systems by using AQNM to consider the effect of ADCs. The result proves that the performance loss caused by low-resolution ADCs can be compensated by increasing the number of BS antennas, which implies the feasibility of installing low-resolution ADCs in massive MIMO systems.

\ifCLASSOPTIONcaptionsoff
  \newpage
\fi


\begin{thebibliography}{1}

\footnotesize{\bibitem{Ref 1}
  J.~G. Andrews, S.~Buzzi, C.~Wan, S. V.~Hanly, A.~Lozano, A. C. K.~Soong, J. C.~Zhang, ``What will 5G be?''  \emph{IEEE J. Sel. Areas Commun.}, vol. 32, no. 6, pp. 1065-1082, Jun. 2014.}
\bibitem{Ref 2}
  T. L. Marzetta, ``Noncooperative cellular wireless with unlimited numbers of base station antennas,'' \emph{IEEE Trans. Wireless Commun.}, vol. 9, no. 11, pp. 3590-3600, Nov. 2010.
\bibitem{Ref 3}
  H. Q. Ngo, E. G. Larsson, and T. L. Marzetta, ``Energy and spectral efficiency of very large multiuser MIMO systems,'' \emph{IEEE Trans. Commun.}, vol. 61, no. 4, pp. 1436-1449, Apr. 2013.
\bibitem{Ref 4}
  J. Singh, O. Dabeer, and U. Madhow, ``On the limits of communication with low-precision analog-to-digital conversion at the receiver,'' \emph{IEEE Trans. Commun.}, vol. 57, no. 12, pp. 3629-3639, Dec. 2009.
\bibitem{Ref 5}
  Q. Bai, A. Mezghani, and J. A. Nossek, ``On the optimization of ADC resolution in multi-antenna systems,'' in \emph{Proc. of the  Tenth Int. Symposium on Wireless Commun. Systems (ISWCS 2013)}, pp. 1-5, Aug. 2013.

\bibitem{Ref 6}
  O. Orhan, E. Erkip, and S. Rangan, ``Low power analog-to-digital conversion in millimeter wave systems: Impact of resolution and bandwidth on performance,'' arXiv preprint arXiv: 1502.01980, available in \url{http://arxiv.org/abs/1502.01980}.

\bibitem{Ref 7}
  Q. Zhang, S. Jin, K. K. Wong, H. B. Zhu, and M. Matthaiou, ``Power scaling of uplink massive MIMO systems with arbitrary-rank channel means,'' \emph{IEEE J. Sel. Topics Signal Process.}, vol. 8, no. 5, pp. 966-981, Oct. 2014.



\bibitem{Ref BaiQ}
Q. Bai and J. A. Nossek, ``Energy efficiency maximization for 5G multi-antenna receivers,'' \emph{Trans. Emerging Tel. Tech.}, vol. 26, no. 1, pp. 3-14, Jan. 2015.
\bibitem{Ref 10}
  R. Krishnan, M. Khanzadi, N. Krishnan, Y. Wu, A. A. I. Graell, T. Eriksson, and R. Schober, ``Linear massive MIMO precoders in the presence of phase noise: A Large-Scale analysis,'' \emph{IEEE Trans. Veh. Tech.}, vol. PP, no. 99, pp. 1-1, Jan. 2015.


\bibitem{Ref 9}
  B. Hassibi and B. M. Hochwald, ``How much training in needed in multiple-antenna wireless link?'' \emph{IEEE Trans. Inf. Theory}, vol. 49, no. 4, pp. 951-963, Apr. 2003.

\end{thebibliography}
\end{document}